\documentclass[aps,pra,twocolumn,floatfix,floats,showpacs,superscriptaddress,raggedbottom]{revtex4-1}
\usepackage{graphicx,latexsym}
\usepackage{dcolumn}
\usepackage{amsmath}
\usepackage{amssymb,bm}
\usepackage[normalem]{ulem}

\def\mb#1{\mbox{\boldmath$#1$}}

\def\eq#1{Eq.\ (\ref{#1})}
\def\fig#1{Fig.\ \ref{#1}}

\usepackage{hyperref}
\hypersetup{
    pdfnewwindow=true,       
    colorlinks=true,         
    linkcolor=blue,          
    citecolor=blue,          
    filecolor=magenta,       
    urlcolor=black           
}

\usepackage{booktabs}

\begin{document}


\title{Coherent transient transport of interacting electrons\\ through a quantum waveguide switch}


\author{Nzar Rauf Abdullah}
\email{nra1@hi.is}
\affiliation{Science Institute, University of Iceland,
        Dunhaga 3, IS-107 Reykjavik, Iceland}

\author{Chi-Shung Tang}
 \affiliation{Department of Mechanical Engineering,
        National United University, 1, Lienda, Miaoli 36003, Taiwan}

\author{Andrei Manolescu}
 \affiliation{Reykjavik University, School of Science and Engineering,
              Menntavegur 1, IS-101 Reykjavik, Iceland}

\author{Vidar Gudmundsson}
\email{vidar@hi.is}
 \affiliation{Science Institute, University of Iceland,
        Dunhaga 3, IS-107 Reykjavik, Iceland}

%

\begin{abstract}
We investigate coherent electron-switching transport in a double
quantum waveguide system in a perpendicular static or vanishing magnetic field. The
finite symmetric double waveguide is connected to two semi-infinite
leads from both ends. The double waveguide can be defined as two
parallel finite quantum wires or waveguides coupled via a window to facilitate
coherent electron inter-wire transport. By tuning the length of the
coupling window, we observe oscillations in
the net charge current and a maximum electron conductance for
the energy levels of the two waveguides in resonance. 
The importance of the mutual Coulomb interaction
between the electrons and the influence of two-electron states is
clarified by comparing results with and without the interaction. 
Even though the Coulomb interaction can lift two-electron states out of
the group of active transport states the length of the coupling window
can be tuned to locate two very distinct transport modes in the system
in the late transient regime before the onset of a steady state. 
A static external magnetic field and quantum-dots formed by side gates 
(side quantum dots) can be used to enhance the inter-waveguide transport 
which can serve to implement a quantum logic device. The fact that the 
device can be operated in the transient regime can be used to enhance its
speed. 

\end{abstract}



\maketitle


\section{Introduction}

Various schemes associated with quantum computing
have been proposed for quantum information storage and transfers,
such as superconducting coplanar waveguide resonators and dual
waveguide devices~\cite{PhysRevB.89.115417,Ionicioiu.15.125}. A
waveguide can be defined as a quantum wire in which the electron
wave propagates in quantized modes without losing phase coherence at
low temperature.  Various proposals have been suggested to
implement a semiconductor qubit~\cite{PhysRevLett.89.117901,
SemicondSciTechnol.19.412, PhysRevA.72.032330}. Among these, an idea
is to use parallel quantum waveguides with a coupling-window placed
between them~\cite{ApplPhysLett.81.22}. Tuning the window-coupling
allows an electron wave interference between the waveguides and can
switch the electron motion from one waveguide to the
other~\cite{ApplPhysLett.79.14}.

The interference in the coupling-region specifies
the possible types of qubit-operations: The electron wave can be
transferred through either waveguide only (pass gate-operation).
It can travel through both waveguides, split equally, implementing the so called
square-root-of-Not ($\sqrt{\rm NOT}$)
operation~\cite{PhysRevA.70.052330}. It can switch totally from one
waveguide to the other
(Not-operation)~\cite{NANOTECHNOLOGYIEEE.1.3}. Or it might switch
from the first waveguide to the second one and then re-enter the first
waveguide (CNot-operation)~\cite{PhysRevLett.84.5912}. 
There are several
parameters and phenomena that can modify the efficiency of the kind of
qubit-operation that are mentioned above such as external magnetic
field, the Coulomb interaction, electrostatic gate voltage, and the geometry of
the coupling window between the waveguides.

Ferry's group proposed magnetically switching transport in an
asymmetric double quantum waveguide, in which an external magnetic
field is applied to switch electron motion from one
quantum-waveguide to the
other~\cite{ApplPhysLett.79.14,PhysRevLett.84.5912}. Another
suggestion for a switching-qubit is to vary the
length of the coupling window (CW) $L_{\rm CW}$. By varying the length
of the CW, a maximum inter-waveguide tunneling can be found that
increases the efficiency of the device~\cite{ApplPhysLett.79.14}.
Another approach for a switching-qubit is to define a saddle
potential in the coupling window instead of a hard-walled well potential. The
saddle-potential makes a soft barrier between the two waveguides
that washes out resonance peaks, but decreases the efficiency of the
qubit~\cite{NANOTECHNOLOGYIEEE.6.5}.

In the present work, we consider a double quantum
waveguide (DQW) system with a CW between the waveguides in a 
static perpendicular magnetic field.  
Both ends of the DQW are connected to semi-infinite
leads with an applied source-drain bias. The coherent electron
transport is investigated in the system by using a non-Markovian
master equation~\cite{1367-2630-11-7-073019,Vidar11.113007,PhysRevB.81.155442,PhysRevB.82.195325}. Smooth
Gaussian potentials are used to define the barrier between the
waveguides and the saddle-like CW potential. In addition, two dots
are embedded in the waveguides close to the CW to enhance inter-waveguide 
transport. We will investigate coherent electron transport in the system by
varying the CW length.

Our results reveal the following conclusions:
First, we predict a maximum conductance at resonant energy-levels.
At a resonance energy, inter-waveguide scattering is enhanced
and the charge density splits equally between the waveguides 
in the absence of the Coulomb interaction as two-electron states
are activated in the transport.
In this case, the DQW essentially works as a $\sqrt{\rm NOT}$-operation
qubit. In the presence of the Coulomb interaction,
the charge current density does not split equally between the waveguides anymore
at the resonance since two-electron states are blocked for the bias range used. 
The strength of the blocking can be varied by the geometry or the material of the system.
The transient transport through two capacitively coupled parallel quantum dots
has been investigated by Moldoveanu et al.\ in a lattice model for a weaker
Coulomb interaction than is assumed here~\cite{PhysRevB.82.085311}.
Second, an external static magnetic field can increase inter-waveguide 
backward scattering leading to a decrease in the efficiency of the qubit. 
These results make clear the difference to previous schemes where variable magnetic
field was used to control the switching between asymmetric 
waveguides~\cite{ApplPhysLett.81.22,NANOTECHNOLOGYIEEE.6.5}.
Third, the influence of the side quantum dots (QDs) can be to
induce more energy-levels into resonance, and consequently
increase the coupling between the waveguides and the inter-waveguide
transport mechanism. In the presence of side QDs, a Not-operation
qubit is realized. The efficiency is slightly decreased in the
presence of side QDs because inter-waveguide backward scattering is
enhanced. At high magnetic field such that the effective magnetic
length is comparable to the radius of the dots, variation of length
of the CW does not influence the electron transport characteristics
anymore because the electrons are localized in the side  dots. Thus,
the net charge current is extremely suppressed.

The paper is organized as follows: In Sec.~\ref{Sec:II}, we present
the model describing the window-coupled double waveguide system based on the
quantum master equation (QME)
approach. Section \ref{Sec:III} presents our numerical results and
discussion. Concluding remarks are addressed in Sec.~\ref{Sec:IV}.

\section{Model and Theory}\label{Sec:II}

\begin{figure}[htbq!]
 \includegraphics[width=0.35\textwidth,angle=0]{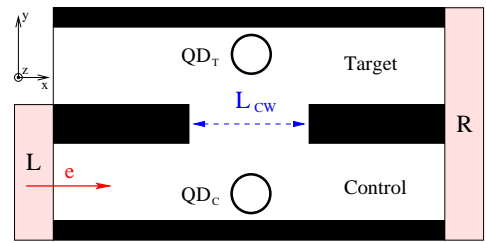}
 \caption{(Color online)
      Schematic diagram depicting the double waveguide with a CW of length $L_{\rm CW}$
      (blue dashed arrow). A quantum dot QD$_{\rm C}$ (QD$_{\rm T}$)
      is embedded in the control (target) waveguide.
      An electron from left side enters the control waveguide (red arrow).
      Asymmetric contacts are indicated by the light red 
      rectangles marked L and R.}
      \label{fig01}
\end{figure}

In this section, we present a method for calculating the
transient ballistic transport in the coupled-window quantum
waveguides without and with side QDs. We consider two laterally
coupled waveguides connected to two semi-infinite wires from both
ends. The lower waveguide (control) and the upper waveguide (target)
are coupled through a CW with length $L_{\rm CW}$ as is
schematically shown in \fig{fig01}. We consider
a control-QD (QD$_{\rm C}$) and a target-QD (QD$_{\rm T}$) embedded,
respectively, in the control- and the target-waveguide near the CW. The
total system is exposed to an external magnetic field in the
$z$-direction B${\rm \hat{z}}$. The system is designed such that the
electrons are injected from the left lead to only the
control-waveguide (red arrow).

Our aim here is to study the charging of, and the output from each
of the waveguides by varying the CW length $L_{\rm CW}$ (blue dashed
arrow). In addition, the effects of the Coulomb interaction, 
the external magnetic field, and the side dots on the conductance will be explored.

The DQW is a two dimensional system with hard-wall confinement in
the contact area to the external leads at $x= \pm L_x/2$ with $L_x$
the length of the waveguides in the transport direction and parabolic
confinement in the $y$-direction. The single-electron Hamiltonian of
the DQW including side QDs in an external magnetic field is
\begin{equation}
       h_\mathrm{S}= \frac{(\mathbf{\mb{p}} + e\mathbf{A}_{\rm ext})^2}{2m^*} + V_{\rm c}(x,y)  +
       V_{\rm DW}(\mathbf{r}) + V_{\rm QDs}(\mathbf{r})\, ,
\label{h_S}
\end{equation}
where $\mb{p}$ indicates the canonical momentum, ${\bf A}_{\rm ext}
= (0,-By, 0)$ is vector potential of the external magnetic field,
and $m^*$ is the effective mass of an electron. The confining
potential of the DQW is $V_{\rm c}(x,y) = V_{\rm c}(x) + V_{\rm
c}(y)$, where $V_{\rm c}(x)$ stands for a hard-wall confining
potential and $V_{\rm c}(y)=\frac{1}{2} m^*{\Omega^2_\mathrm{0}}y^2$
refers to a parabolic confining potential with characteristic energy
$\hbar\Omega_\mathrm{0}$. The Gaussian potential barrier between the
two waveguides is presented by $V_{\rm DW}(\mathbf{r})$ and the side
QDs are defined by $V_{\rm QDs}(\mathbf{r})$. We diagonalize the
Hamiltonian of the central system presented in \eq{h_S} using a
basis~\cite{PhysRevB.82.195325} built with the eigenfunctions of the first two terms of 
(\ref{h_S}) to find the
single-electron energy spectrum $E_{\rm n}$, where $n$ is the
composite quantum number of a single-electron state.

Below, we shall demonstrate how the DQW is connected to the leads and show
the time evolution of electron in the system.

\subsection{DQW connected to leads}

We connect the DQW to two leads that act as electron reservoirs via
a contact region. The central system and the leads have the same
confinement energy and the same strength of the perpendicular
magnetic field. The total Hamiltonian of the system is then
\begin{align}
      H(t) &= \sum_n E_n d^\dagger_n d_n + \frac{1}{2}\sum_{ijrs}\langle V_{\textrm{Coul}}\rangle
        d_i^{\dagger}d_j^{\dagger}d_sd_r \nonumber \\
           &+ \sum_{l=\mathrm{L,R}}  \int d{\mb{q}}\, \epsilon^l(\mb{q}) {c^l_{\mb{q}}}^\dagger c^l_{\mb{q}} \\ 
           &+ \sum_{l=\mathrm{L,R}}  \chi^l(t) \sum_{n}\int d{\mb{q}}\, \left[  {c^l_{\mb{q}}}^\dagger T^l_{\mb{q}n} d_n
            + d^\dagger_n (T^l_{n\mb{q}})^* c^l_{\mb{q}}\right]. \nonumber
\label{Ht}
\end{align}
The first term describes the central system, where $E_n$ is the
single-electron energy of state $|n\rangle$ of the central system,
and $d^\dagger_n$ $(d_n)$ denotes the electron creation
(annihilation) operator respectively. The second term is the 
mutual Coulomb interaction between electrons in the central system.    
The third term of the
Hamiltonian denotes the semi-infinite $l^{\mathrm{th}}$ lead, where $l$ refers to
the left L or the right R lead with the dummy index
$\mb{q}$ representing the ``momentum'' of an electron in the lead and
its subband index~\cite{PhysRevB.82.195325}, $\epsilon^l(\mb{q})$ is the
single-electron energy spectrum in lead $l$, and the electron
creation and annihilation operator in lead $l$ are
${c^l_{\mb{q}}}^\dagger$ and $c^l_{\mb{q}}$, respectively.

The last term is the time-dependent coupling Hamiltonian that connects the central system, the DQW to
the external leads via a coupling tensor $T^l_{\mb{q}n}$, and $\chi^l(t)$ dictates the time-dependence
of the coupling. The coupling Hamiltonian describes
the transfer of an electron between the single-electron state
of the central system $|n\rangle$ and the single-electron state of
the leads $|\mb{q}\rangle$
\begin{equation}
      T^l_{\mb{q}n} =
      \int d\mathbf{r} d\mathbf{r^{\prime}} \psi^l_{\mb{q}}(\mathbf{r}')^*
      g^l_{\mb{q}n} (\mathbf{r},{\bf r'}) \psi^\mathrm{S}_n({\bf r}).
\label{Tlqn}
\end{equation}
$\psi^\mathrm{S}_n({\bf r})$ and $\psi^l_{\mb{q}}(\mathbf{r}')$ are the corresponding single-electron wave
functions of the DQW system and the leads. $g^l_{\mb{q}n} (\mathbf{r},{\bf r'})$
is a nonlocal coupling kernel
\begin{align}
      g^l_{\mb{q}n} ({\bf r},{\bf r'}) &=
                   g_0^l\exp{\left[-\delta_x^l(x-x')^2-\delta_y^l(y-y'-\alpha)^2\right]}
                   \nonumber \\
                   & \times \exp{\left( -\Delta_{n}^l(\mb{q}) / \Delta
                   \right)}, \label{cf}
\end{align}
where $g_0$ is the coupling strength, $\delta_x^l$ and $\delta_y^l$ are the coupling range parameters
in the $x$- and $y$-direction, respectively,
$\Delta$ controls together with $\Delta_{n}^l(\mb{q}) = |E_n-\epsilon^l(\mb{q})|$
the energy affinity between states in the leads and the central system~\cite{Vidar11.113007},
and $\alpha$ is a skewing parameter that shifts the weight of the
coupling to the lower waveguide for an appropriate value stated below.

\subsection{QME and Charge current density}

In this subsection, we present how the time evolution of the open system
is calculated in order to study time-dependent transport properties within
the QME approach~\cite{Breuer2002}.
The time evolution of our system obeys the quantum Liouville-von
Neumann equation
\begin{equation}
      \frac{dW(t)}{dt}= -\frac{i}{\hbar}\left[H(t),W(t)\right],
\end{equation}
where $W(t)$ is the density operator of the total system.

In order to focus our attention on the open central system we
introduce a reduced density operator (RDO) with respect to it, by taking the
trace over the Fock space with respect to the leads $\rho(t)={\rm
Tr}_\mathrm{L} {\rm Tr}_\mathrm{R} W(t)$ with the initial condition
$W(t<t_0)$ = $\rho_\mathrm{L}\rho_\mathrm{R}\rho_\mathrm{S}$, where
$\rho_\mathrm{L}$ and $\rho_\mathrm{R}$ are the density operators of
the isolated left and the right leads, respectively. The density
operator of the $l^\mathrm{th}$ lead can be defined as $\rho_l=
e^{-\beta (H_l-\mu_l N_l)}/{\rm Tr}_l \{e^{-\beta(H_l-\mu_l
N_l)}\}$, where $\beta=1/(k_BT)$ and $k_B$ is the Boltzmann
constant, and $N_l$ is the number of electrons in the leads.
The density operator of the central system at $t =
t_0$ being $\rho(t_0) = \rho_\mathrm{S}$~\cite{Haake3.1723}  allows
us to obtain the equation of motion for the RDO in the following
form~\cite{Haake1973}
\begin{equation}
      \frac{d{\rho}(t)}{dt} = -i{\cal L}_{\rm S}\rho(t) +\int_{t_0}^{t}
      dt' {\cal K}(t,t')\rho (t'),
\end{equation}
where ${\cal L}_{\rm S}\cdot =[H_S,\cdot]/\hbar$ is the Liouvillian with respect to the
time-independent Hamiltonian $H_S$ of the central system and
${\cal K}(t,t')$ is an integral kernel with its roots in the dissipative
time-dependent coupling to the leads~\cite{Haake1973,PhysRevB.82.195325}.

In order to explore the switching processes between the waveguides, we define
the net charge current as
\begin{equation}
      I_{\rm Q}(t) = I_{\rm L}(t) - I_{\rm R}(t),
\label{I_Q}
\end{equation}
where $I_{\rm L}(t)$ refers to the partial current from the left lead into
the control-waveguide and $I_{\rm R}(t)$ indicates the partial current
into the right lead from both waveguides~\cite{Nzar.25.465302}.
In \eq{I_Q} the negative sign in front of $I_{\rm R}$ is since
a positive charge current is defined as the current into the central system,
but the currents $I_L$ and $I_R$ are defined positive from left to right.

To monitor the dynamic evolution of the electrons in the central system,
we calculate the expectation value of charge current density operator
\begin{equation}
      \mathbf{J}({\bf r},t) = {\rm Tr} \left( \hat{\rho}(t)  \mathbf{\hat{J}}({\bf r}) \right),
\end{equation}
where the charge current density operator is defined by
\begin{align} \nonumber
  \mathbf{\hat{J}}({\bf r}) &=
         \sum_{nn'} \Bigg( \frac{e\hbar}{2 m^* i} \Big[ \psi^\mathrm{S*}_{n}({\bf r}) (\nabla \psi^\mathrm{S}_{n'}({\bf r}))
         - (\nabla \psi^\mathrm{S*}_{n}({\bf r})) \psi^\mathrm{S}_{n'}({\bf r})  \Big] \\
         &+\frac{e^2}{m^*} \mathbf{A}_{\rm ext}({\bf r}) \psi^\mathrm{S*}_{n}({\bf r})
         \psi^\mathrm{S}_{n'}({\bf r}) \Bigg) d_n^{\dagger} d_{n'}.
\end{align}

In the following, we shall demonstrate the influence of the length 
of the CW, external
magnetic field, and the side QDs on the coherent electron transport
through the system and the charge current density in the central
system in order to investigate inter-waveguide forward and backward
scattering processes.


\section{Results}\label{Sec:III}

In this section, we shall discuss our numerical results of the
ballistic transport properties through the double waveguide system
made of a GaAs semiconductor with the electron effective mass
$m^*=0.067m_e$. The length of the central system is $L_x=300$~nm
which is assumed to be much smaller than the phase coherent length
$L_{\phi}$. At low temperature $T \sim 0.1 - 2.0$~K, the phase
coherent length of semiconductor based (GaAs/Al$_{\rm 1-x}$Ga$_{\rm
x}$As) electron waveguide is $\sim (30 - 40)\times10^{3}$~nm~\cite{Datta1995}. 
The coherence in the electron
transport in the double waveguide is an essential requirement for
constructing a qubit. For the system placed in a static or vanishing
external magnetic field, its length scales can be
characterized by the effective magnetic length $a_w = \{\hbar /(m^*
\sqrt{(\omega^{2}_c + \Omega^2_0)})\}^{1/2}$, where the cyclotron
frequency $\omega_c = eB/m^*$ and the transverse confinement energy
is $\hbar \Omega_0 = 1.0$~meV. Numerically the effective magnetic
length can be expressed as
\begin{align}
a_w &= \left(\frac{\hbar}{m^*\Omega_0}\right)^{1/2}
    \left(\frac{1}{1+(eB/(m^*\Omega_0))^2}\right)^{1/4}\nonumber\\
    &= \frac{33.74}{ \sqrt[4]{1+2.982[B({\rm T})]^2} }\ {\rm nm}.
\end{align}
For the case of a vanishing magnetic field we assign to it a finite
very small value in order to avoid numerical problems caused
by exact degeneration of states with opposite spin. 

We assume the central system is connected to the external leads
acting as electron reservoirs with chemical potential in the left
(right) lead $\mu_L = 4.0$~meV ($\mu_R = 3.0$~meV), implying an
applied potential difference or bias window $\Delta\mu = eV_{\rm
bias}  = 1.0$~meV. Furthermore, the temperature of the leads is
fixed at ${\rm T}=0.5$~K.

\begin{figure}[htbq!]
 \includegraphics[width=0.48\textwidth,angle=0,bb=60 60 400 264,clip]{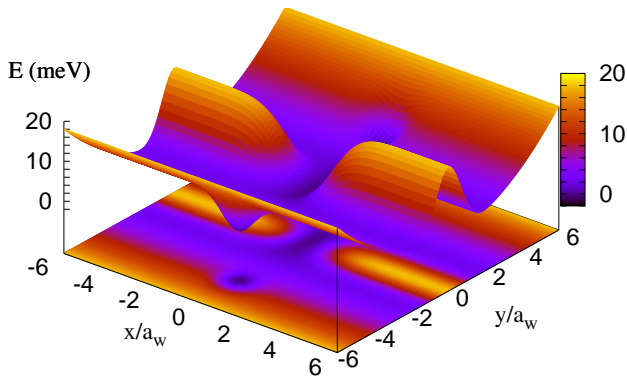}\\
 \caption{(Color online) The potential (in meV) defining the
     DQW central system with the CW and the side QDs. The parameters are
     $\hbar\Omega_0 = 1.0$~meV, $B=0.0$~T, $a_w=33.72$~nm, $L_{\rm CW} = 100$~nm, and $R_{\rm Dot} = 25$~nm.}
     \label{fig02}
\end{figure}

The potential of the double quantum waveguides is shown in
\fig{fig02} with the side QDs located near the CW in order to
influence the inter-waveguide transport.
In the following, we shall explore the influence of the CW and the
side QDs on the charging of and the transport through the system.

\subsection{DQW without side QDs}

We start by considering an ideal window-coupled DQW without side
QDs. The injected electrons are assumed to come from the left lead
only into the control-waveguide. This asymmetry in the injection is
accomplished with the choice of the skewing parameter $\alpha =
4.0a_w$ defined in Eq.\ (\ref{cf}) and only used for the left lead.  
In addition, we assume the two quantum waveguides to be
of the same width.  They are separated by a Gaussian potential barrier, in
which the control- and target-waveguide interact with each other via
a CW. The DQW system is described by
\begin{equation}
      V_{\rm DQW}(\mathbf{r}) = V_{\rm B}(y) + V_{\rm CW}(x,y),
\end{equation}
where  $V_{\rm B}(y)$ is the barrier potential between the quantum waveguides
defined as
\begin{equation}
      V_{\rm B}(y) = V_0\; {\rm exp}{(-\beta_0^2y^2)},
\end{equation}
with $V_0=18.0$~meV and $\beta_{0}=0.03$ nm$^{-1}$, as well as a CW
potential
\begin{equation}
      V_{\rm CW}(x,y) = -V_0\; {\rm exp}{(-\beta_x^2 x^2-\beta_y^2 y^2)},
\end{equation}
with $V_0=18.0$~meV, and $\beta_y=0.03$ nm$^{-1}$ implying a barrier
width $W_{\rm B} \simeq 66.5$ nm for the first subband which can
prevent electron tunneling between the waveguides through the
barrier. $\beta_x$ is a parameter that determines the CW length.
Thus, the length of the CW can be defined as $L_{\rm CW} =
2/\beta_x$.

\begin{figure}[htbq!]
 \includegraphics[width=0.23\textwidth,angle=0,bb=54 50 211 294,clip]{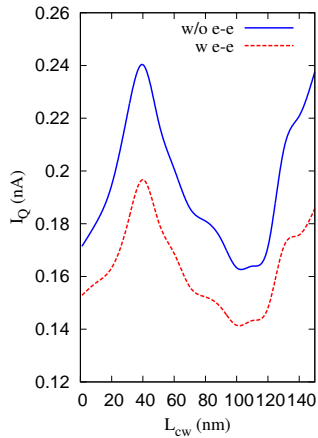}
 \caption{(Color online) The net charge current $I_{\rm Q}$ versus CW length $L_{\rm CW}$
          without (w/o) (blue solid) and with (w) (red dashed) electron-electron (e-e) interaction
          for magnetic field $B = 0.0$~T
          The chemical potentials are $\mu_L = 4.0$~meV and $\mu_R = 3.0$~meV
          implying $\Delta \mu = 1.0~{\rm meV}$.}
\label{fig03}
\end{figure}

We begin to analyze the results by presenting \fig{fig03} which
shows the net charge current $I_{\rm Q}$ as a function of the CW
length $L_{\rm CW}$ at $t =200$~ps and $B = 0.0$~T for noninteracting
electrons (solid blue) and for Coulomb interacting electrons (dashed red).
The Coulomb interaction reduces the overall current slightly.

The oscillations in the net charge current are indicative of a possible charge transport
between the control-waveguide and the target-waveguide. The
oscillations give rise to a peak and a dip in the net charge
current at $L_{\rm CW} \simeq 40$~ and $110$~nm, respectively.
Similar oscillation features have been found
by Zibold {\it et al}.\ and Gong {\it et al}. in the absence of the Coulomb interaction,
and put in relation to energy-dependent inter-waveguide transmission in the quantum
regime. They pointed out that the inter-waveguide transmission can
be enhanced when the energy levels of the coupled-waveguide system
achieve a resonance conditions for a specific CW
length~\cite{PhysRevB.76.195301,ChinPhysLett.24.8}.

\begin{figure}[htbq!]
  \includegraphics[width=0.23\textwidth,angle=0,bb=54 50 211 294,clip]{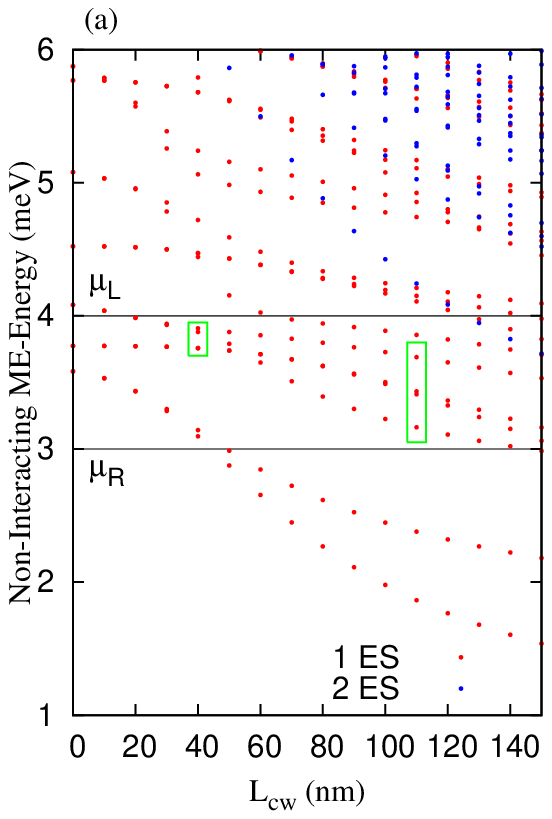}
  \includegraphics[width=0.23\textwidth,angle=0,bb=54 50 211 294,clip]{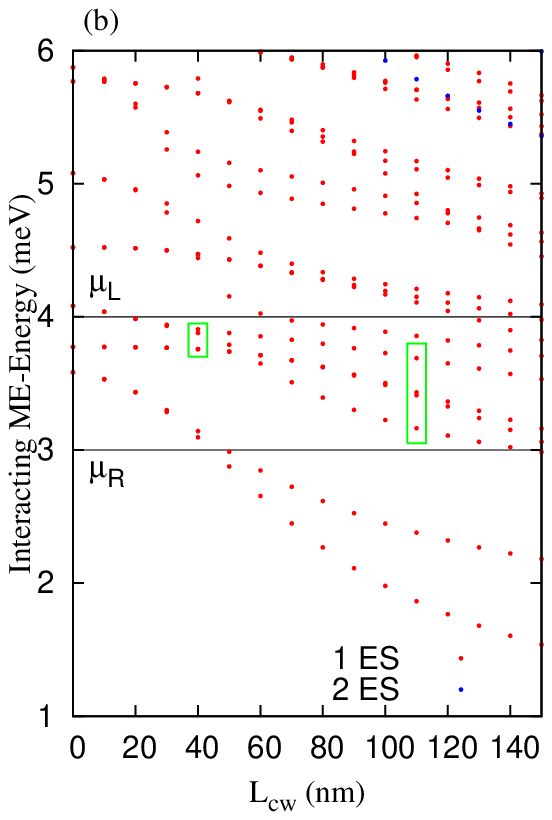}
 \caption{(Color online) ME energy spectrum in the central system as a function of
          CW length $L_{\rm CW}$ without (a), and with (b) the electron-electron Coulomb interaction
          including one-electron states (1ES, red dots) and
          two-electron states (2ES, blue dots) at $B = 0.0$~T. The one-electron states in the left green rectangle
          are close to be in resonance, but the one-electron states in the right green rectangle are not.
          The chemical potentials are $\mu_L = 4.0\ {\rm meV}$ and $\mu_R = 3.0\ {\rm meV}$ (black)
          implying $\Delta \mu = 1.0~{\rm meV}$.}
\label{fig04}
\end{figure}

To demonstrate the resonant energy levels in our system, 
we present \fig{fig04} which shows the many-electron (ME) energy spectra for the DQW
system as a function of the CW length at $B = 0.0$~T including one-electron
states (1ES, red dots) and two-electron states (2ES, blue dots).
The ME spectrum without the mutual Coulomb interaction is displayed
in Fig.\ \ref{fig04}(a), but the interacting spectrum in Fig.\ \ref{fig04}(b).  
The only difference between the energy spectra is that the Coulomb
interaction has raised most of the 2ES well above the bias window in
Fig.\ \ref{fig04}(b). The 2ES do not contribute to the electron transport in the 
presence of the Coulomb interaction because 
they are far from the chosen bias window, leading to a decrease in the net charge current.

In the case of no coupling window the two waveguides are only coupled
through tunneling and the Coulomb interaction~\cite{PhysRevB.82.085311} 
leading to nearly degenerate states. When the length of the CW
$L_{\rm CW}$ is increased we thus find: First, the energy spectrum decreases
monotonically~\cite{PhysRevB.76.195301}. Second, the near degeneration of
the energy-levels is lifted. As a result, this leads to a crossover of
energy levels (resonance) between the control- and target-waveguide
at $L_{\rm CW}\simeq 40$~nm (left green rectangular) which
corresponds to the current-peak in \fig{fig03}. The resonance
between the waveguides enhances the inter-waveguide electron
transport and increases the net charge current. The splitting of
the energy levels increases at a higher window-coupling length such
as $L_{\rm CW}\simeq 110$~nm (right green rectangular) indicating
weaker resonances and more back-scattering inter-waveguide
transport.

\begin{figure*}[htbq]
       \begin{center}
       \includegraphics[width=0.34\textwidth,angle=0,bb=89 59 285 219]{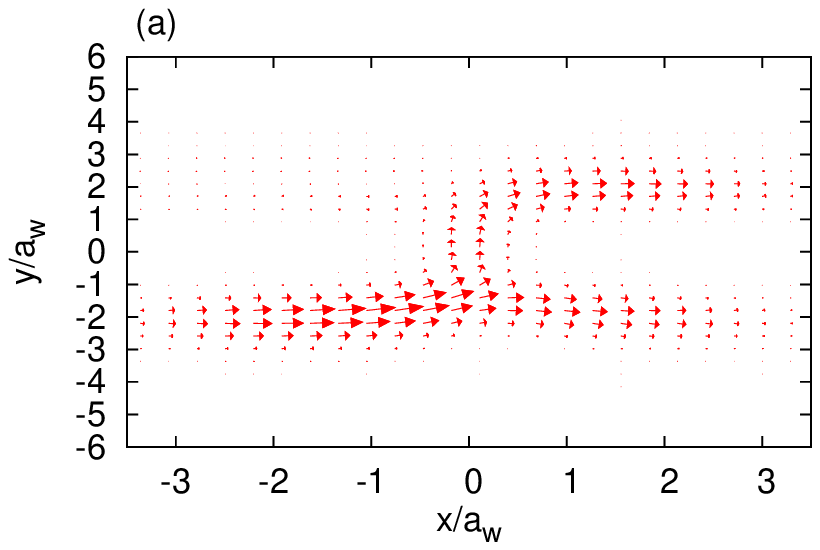}
       \includegraphics[width=0.34\textwidth,angle=0,bb=56 59 252 219]{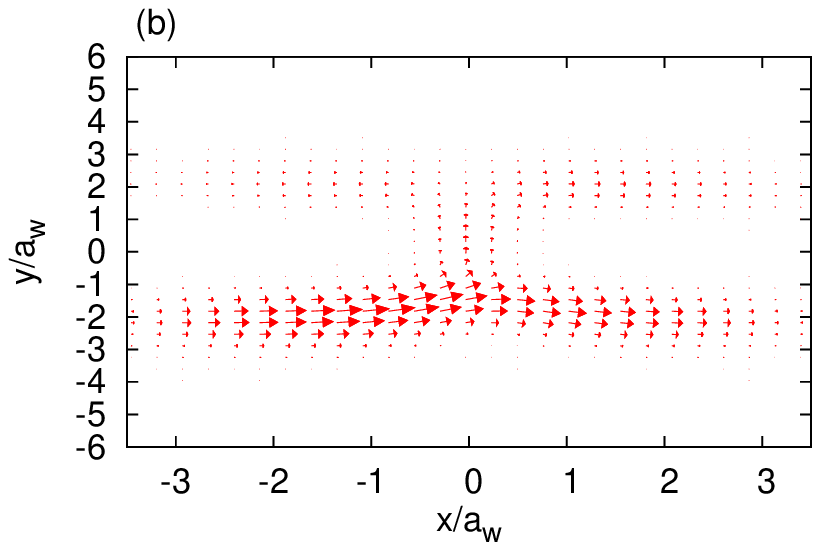}\\  
       \includegraphics[width=0.34\textwidth,angle=0,bb=89 59 285 219]{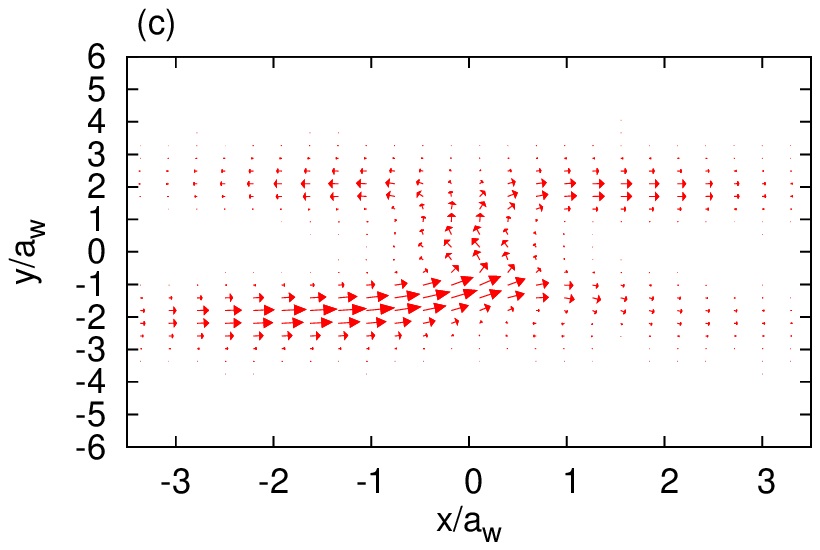}
       \includegraphics[width=0.34\textwidth,angle=0,bb=56 59 252 219]{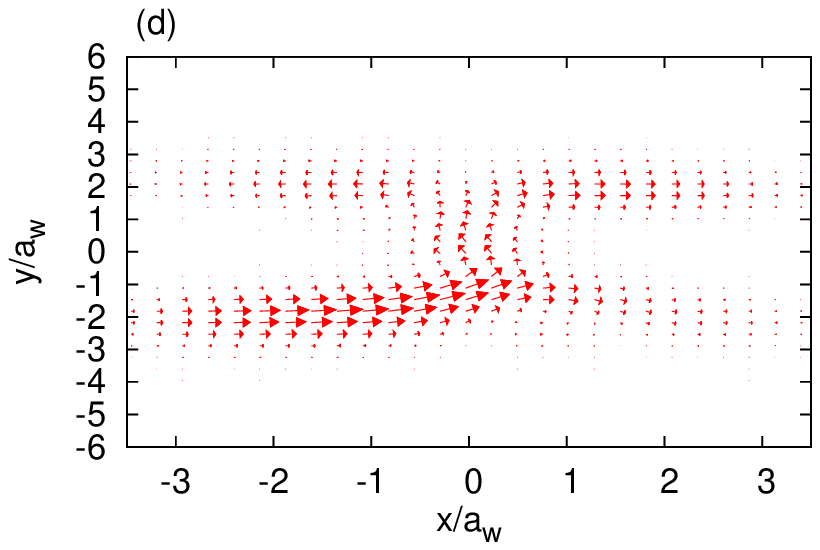}\\    
       \end{center}
       \caption{(Color online) Charge current density in the DQW at $t = 200$~ps
               without (left panels) with (right panels) electron-electron Coulomb interaction
               in the peak and the dip net charge current at $L_{\rm CW}\simeq 40$~nm (a)-(b) and
               $110$~nm (c)-(d), respectively, shown in \fig{fig03}
               in the case of $B = 0.0$~T (blue solid).
               The effective magnetic length is $a_w = 33.72$~nm}
       \label{fig05}
\end{figure*}

To further explain the peak and the dip in the net charge
current, we present the charge current density at $B = 0.0$~T and
$t = 200$~ps in \fig{fig05} for the noninteracting system (left panels) 
and the interacting one (right panels) for the peak and the dip net charge
current at $L_{\rm CW}\simeq 40$~nm (a-b) and $110$~nm (c-d),
respectively, shown in \fig{fig03}.

The dynamical evolution occurring in the CW implements different types of
quantum logic gates in our system.
In \fig{fig05}(a) the charge current density in the absence of the Coulomb interaction
reveals the following electron motion in the waveguides at $L_{\rm CW}\simeq 40$~nm: 
The charge is injected into the control input, then it exits equally 
from both control and target output. In the DQW the charge exhibits
partial inter-waveguide forward scattering which is similar to the
condition of having equal electron transmission ratio in the double
waveguides at a resonance~\cite{PhysRevB.76.195301}. The DQW here
can be defined as a beam splitter or a $\sqrt{\rm NOT}$-operation
quantum logic gate ($\sqrt{\rm NOT}$-operation
qubit)~\cite{PhysRevA.70.052330}.

Figure \ref{fig05}(c) shows the charge current density in the
current-dip at $L_{\rm CW}\simeq 110$~nm. The charge current density
switches from the control- to the target-waveguide with a small
ratio of inter-waveguide backward scattering. The charge switching
property of the system here can be introduced as a Not-operation
quantum logic gate~\cite{NANOTECHNOLOGYIEEE.6.5}. The net charge
current is suppressed here due to the presence of an inter-waveguide
backward scattering that forms the current-dip. The cause of the
inter-waveguide backward scattering is the geometry of the system,
the symmetric waveguides, while in the asymmetric system the back
scattering can be avoided~\cite{ApplPhysLett.81.22}.

With the Coulomb interaction the charge current density is very 
similar for the dip near 110~nm (comparing Fig's \ref{fig05}(c)
and (d)), but different for the peak at 40 nm, where very little
scattering into the target waveguide is seen (comparing Fig. \ref{fig05}(a) and (b)). 
Without the Coulomb interaction 2ES contribute up to 1/3 of the 
charge current, but the interaction reduces this with
Coulomb blocking to a negligible quantity. The forward inter-waveguide
scattering in Fig.\ \ref{fig05}(a) is facilitated by two-electron
processes and states. In the present system the Coulomb interaction
blocks these processes to a large extent.

\begin{figure}[htbq!]
 \includegraphics[width=0.23\textwidth,angle=0,bb=54 50 211 294,clip]{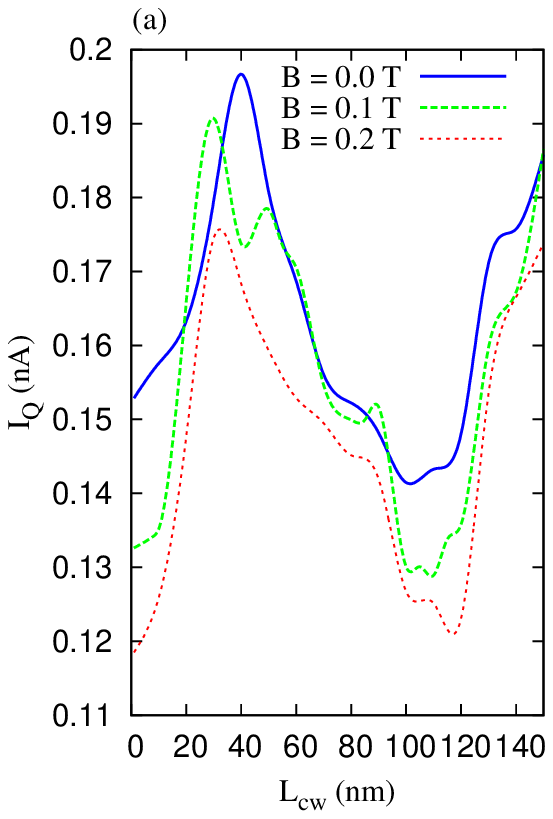}
 \includegraphics[width=0.23\textwidth,angle=0,bb=54 50 211 294,clip]{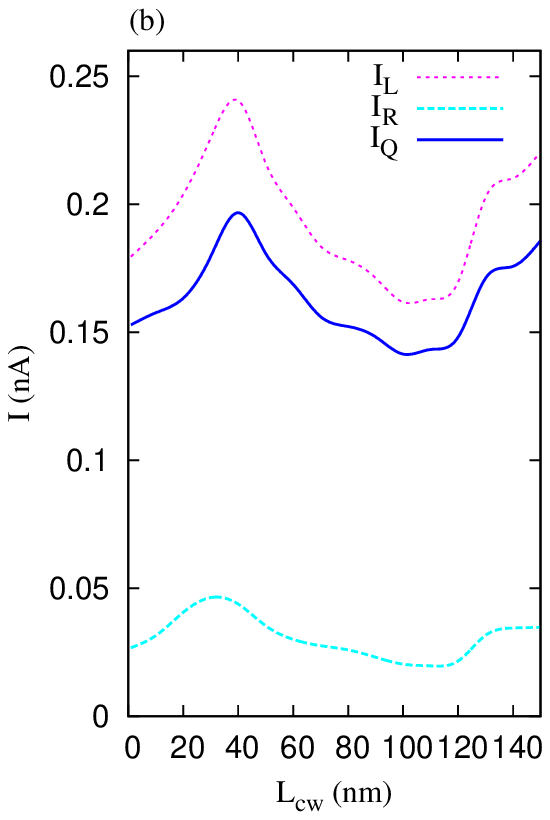}
 \caption{(Color online) (a) The net charge current $I_{\rm Q}$ versus CW length $L_{\rm CW}$ 
          in the presence of the Coulomb interaction
          for different values of the magnetic field $B = 0.0$~T (blue solid), $0.1$~T (green dashed), 
          and $0.2$~T (red dotted)
          at time $t = 200$~ps. (b) The current from the left lead $I_{\rm L}$ (pink dotted),
          the current into the right lead $I_{\rm R}$ (blue dashed), and the net charge current
          $I_{\rm Q}$ (blue solid) for $B = 0.0$~T versus the CW length.
          The chemical potentials are $\mu_L = 4.0$~meV and $\mu_R = 3.0$~meV
          implying $\Delta \mu = 1.0~{\rm meV}$.}
\label{fig06}
\end{figure}

The effects of a static external magnetic field are presented in \fig{fig06}(a) which
shows the net charge current $I_{\rm Q}$ as a function of the CW
length $L_{\rm CW}$ at $t =200$~ps in the presence of the Coulomb interaction
for different values of the magnetic field $B = 0.0$~T (blue solid), $0.1$~T (green dashed),
and $0.2$~T (red dashed). For comparison we display in
\fig{fig06}(b) the left and the right currents $I_{\rm L}$ and
$I_{\rm R}$ at $B = 0.0$~T and the same point in time. The comparison shows that at
$t =200$~ps the system is still in the charging phase, but the
overall shape of the currents is similar.

The role of the external magnetic field in our system is different
from the role it plays in many models where researchers have applied 
a variable magnetic field to switch between different processes in the waveguides. 
The external magnetic field has been used to switch an electron current
from one quantum-waveguide to the other one in a window coupled double
waveguide~\cite{ApplPhysLett.81.22}. In our model, we observe switching
characteristic in {\sl absence} of a magnetic field $B = 0.0$~T. 
In the presence of stronger magnetic
field, the electron motion is affected by the Lorentz force in which
the electrons tend to a circular motion in both control- and
target-waveguides increasing the electron dwell-time in the system.
\begin{figure}[htbq!]
 \includegraphics[width=0.45\textwidth,angle=0]{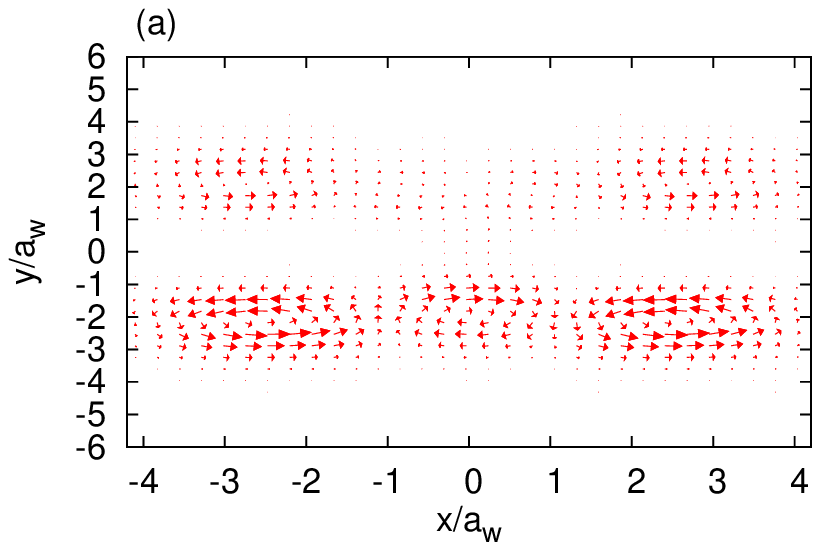}\\
 \includegraphics[width=0.45\textwidth,angle=0]{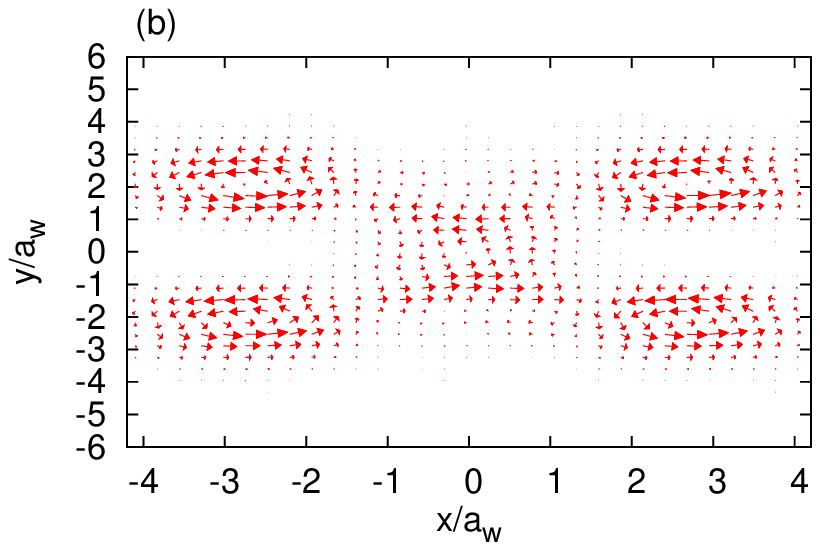}
 \caption{(Color online) Charge current density in the DQW at $t = 200$~ps
 in the presence of the Coulomb interaction
in the peak at $L_{\rm CW}\simeq 40$~nm (a), and the dip at $L_{\rm CW}\simeq 110$~nm (b)
net charge current of \fig{fig06}(a) in the case of $B = 0.2$~T (red dotted).
The effective magnetic length is $a_w = 32.76$~nm.}
\label{fig07}
\end{figure}
The inter-waveguide backward scattering is enhanced while
a suppression in the inter-waveguide forward scattering
is observed in the presence of strong magnetic field.
Therefore, the net charge current decreases as shown in \fig{fig06}(a)
in the case of $B = 0.1$~T (green dashed) and $0.2$~T (red dotted).

To understand the reasons for the suppression of the net charge
current in the presence of a higher magnetic field, we show in
\fig{fig07} the charge current density at magnetic field $B = 0.2$~T
in the peak at $L_{\rm CW}\simeq 40$~nm (a) and the dip at $L_{\rm
CW}\simeq 110$~nm (b) of the net charge current shown in
\fig{fig06}(a) (red dotted). An obvious explanation is: The
perpendicular magnetic field reduces the effective barrier height
and increases the splitting of the energy levels at the crossover
energy or resonance energy (not shown). Therefore the charge current
density does not equally split between the waveguides or localize in only
control-waveguide, but rather an inter-waveguide backward
scattering is observed due to the Lorentz force, consequently the
net charge current is suppressed and we observe edge states
forming in each of the waveguides. In this case the double-waveguide
system does neither play the role of a $\sqrt{\rm NOT}$-operation nor
a Not-operation qubit in the presence of a higher magnetic field.

In the following, we shall study the influence of the side QDs in
the DQW for different values of the magnetic field.

\subsection{DQW with side QDs}

In this section, we consider a side dot embedded in each waveguide
to enhance the inter-waveguide transport as is shown in \fig{fig02}.
The dots to the sides of the CW may be expected to increase the
dwell time in the coupling region and change the resonance condition
between the two waveguides. The window-coupling DQW potential with
side QDs is described by
\begin{equation}
      V_{\rm DQW}(\mathbf{r}) = V_{\rm B}(y) + V_{\rm CW}(x,y) + V_{\rm Dot}(x,y),
\end{equation}
where $V_{\rm Dot}(x,y)$ is the side dot potential that is defined as
\begin{equation}
      V_{\rm Dot}(x,y) = -V_{\rm D}\; {\rm exp}{(-\gamma_x^2 x^2-\gamma_y^2 (y-y_0)^2)},
\end{equation}
with $V_{\rm D} = 8.0$~meV, $y_0 = 5.6$a$_w$, and $\gamma_{\rm x} = \gamma_{\rm y} = 0.04$ nm$^{-1}$ implying
the radius of each side-dot $R_{\rm Dot} \approx 25$~nm.

\begin{figure}[htbq!]
 \includegraphics[width=0.23\textwidth,angle=0,bb=54 50 211 294,clip]{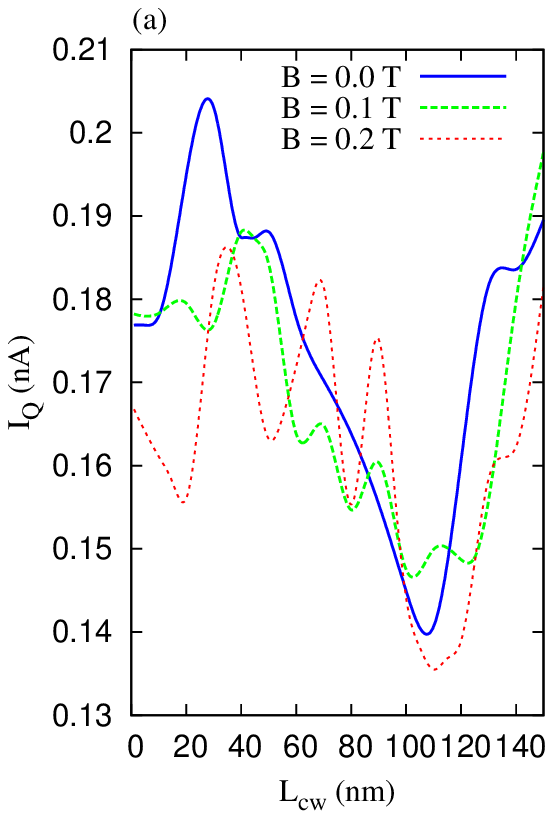}
 \includegraphics[width=0.23\textwidth,angle=0,bb=54 50 211 294,clip]{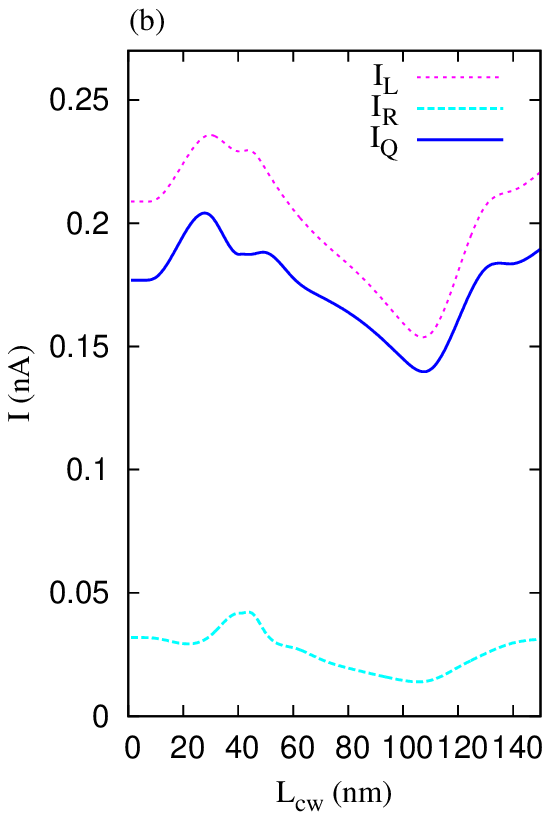}
 \caption{(Color online) The net charge current $I_{\rm Q}$ versus the window-coupling length $L_{\rm CW}$
          in the presence of the Coulomb interaction
          for different magnetic field $B = 0.0$~T (blue solid), $0.1$~T (green dashed),
          $0.2$~T (red dotted) ,$0.9$~T (pink dotted)
          at time $t = 200$~ps. (b) The current from the left lead $I_{\rm L}$ (pink dotted),
          the current into the right lead $I_{\rm R}$ (blue dashed), and the net charge current
          $I_{\rm Q}$ (blue solid) for $B = 0.0$~T versus the CW length.
          The chemical potentials are $\mu_L = 4.0$~meV and
          $\mu_R = 3.0$~meV implying $\Delta \mu = 1.0~{\rm meV}$.}
\label{fig08}
\end{figure}

Figure \ref{fig08}(a) shows the net charge current as a function
of the CW length at $t = 200$~ps in the presence of the Coulomb interaction 
for different values of the magnetic field $B = 0.0$~T (blue solid), $0.1$~T (green dashed), and
$0.2$~T (red dashed). An oscillation in
the net charge current is again established in a vanishing magnetic field
$0.0$~T, while at higher magnetic field values ~$0.1$ and $0.2$~T
some extra fluctuations in the current oscillation characteristic are observed.
The existence of fluctuations correlates with dynamic motion of the charge in the system
in which electrons participate in an inter-waveguide backward or forward scattering at different
CW length. If the magnetic field is increased to $B = 0.9$~T, such that 
the effective magnetic length is close to the dot radius $a_w = R_{\rm
Dot}$ the net charge current is suppressed
to vanishing values (not shown). In that case, varying of the CW length
does not affect the electron transport in the system. There are
several reasons that lead to the almost vanishing net charge
current at high magnetic field such as edge effects, inter-waveguide
backward scattering and localized electrons in the dots as the
effective magnetic length is close to the dot radius.

We present \fig{fig08}(b) to compare the left partial
current (pink dotted), the right partial current (blue dashed) to the
net charge current (blue solid) at $B = 0.0$~T. It can be seen that the 
partial currents have the same oscillation features and develop a peak 
and a current dip at the same CW lengths.

\begin{figure}[htbq!]
\includegraphics[width=0.35\textwidth,angle=0]{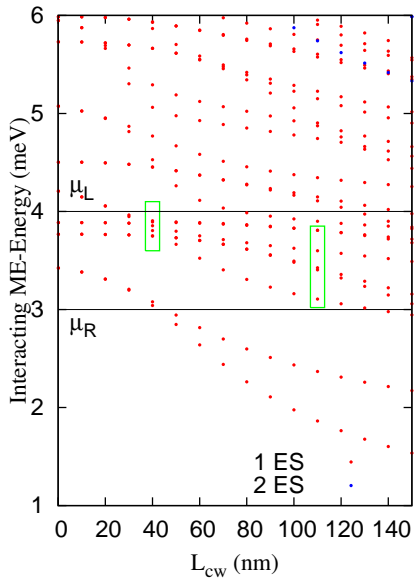}\\
\caption{(Color online) The interacting ME energy spectrum of the central system
          with side QDs as a function of
          the CW length $L_{\rm CW}$ including one-electron states (1ES, red dots) and
          two electron states (2ES, blue dots) at $B = 0.0$~T. The one-electron states in the left green rectangular
          are close to be in resonance, but the one-electron states in the right green rectangular are not.
          The chemical potentials are $\mu_L = 4.0\ {\rm meV}$ and $\mu_R = 3.0\ {\rm meV}$ (black)
          implying $\Delta \mu = 1.0~{\rm meV}$.}
\label{fig09}
\end{figure}

In order to understand the characteristics of net charge current,
we refer to the energy spectrum and the charge current density.
Figure \ref{fig09} shows the ME-energy versus the window-coupling
length at $B = 0.0$~T including one-electron states (1ES, red dots) and two
electron states (2ES, blue dots). 
It should be noted that more energy levels get into resonance at
$L_{\rm CW} \simeq 40$~nm (left green rectangular) in the
presence of the side-gate dots which indicates a stronger coupling
between the control- and the target-waveguide. At larger CW length
such as $L_{\rm CW}\simeq 110$~nm (right green rectangular), the
energy splitting increases resulting in different electron motion in
the system.

\begin{figure}[htbq!]
      \includegraphics[width=0.45\textwidth,angle=0]{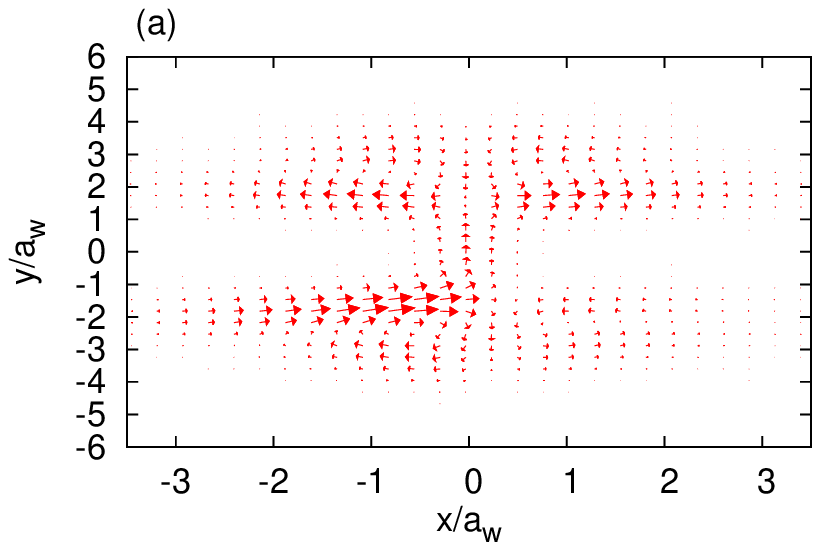}\\
      \includegraphics[width=0.45\textwidth,angle=0]{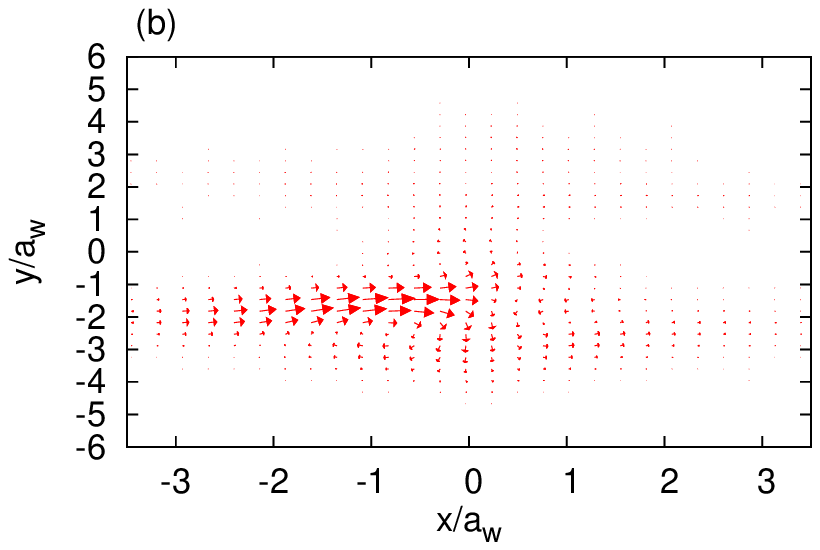}
      \caption{(Color online) Charge current density in the DQW at $t = 200$~ps
               in the presence of the Coulomb interaction
               in the peak $L_{\rm CW} \simeq40$~nm (a), and the dip $L_{\rm CW} \simeq 110$~nm (b)
               of the net charge current of \fig{fig08}(a) in the case of $B = 0.0$~T.
               Other parameters are $a_w = 33.72$~nm, and $R_{\rm Dot} \approx 25$~nm.}
\label{fig10}
\end{figure}

The electron motion in the waveguides is completely changed in the
presence of the side QDs. Figure \ref{fig10} displays the charge
current density at $B = 0.0$~T and $L_{\rm CW} \simeq 40$~nm (a) and current-dip
at $L_{\rm CW} \simeq 110$~nm (b) shown in \fig{fig08}(a) (blue solid).
In \fig{fig10}(a) the charge current density is displayed for 
$L_{\rm CW} \simeq 40$~nm. The charge in the
control-waveguide is switched to the target-waveguide, thus a
Not-operation is realized. Comparing to the system without side-gate
dots in \fig{fig05}(b), the inter-waveguide transport is enhanced
due to a stronger coupling between the waveguides at the
resonant-energy levels at $L_{\rm CW} \simeq 40$~nm. 
Therefore, the side QDs play an essential role
to convert the qubit operation in the system. In addition, the
inter-waveguide backward scattering here reduces the efficiency of
the qubit. As we mention previously, the existence of
inter-waveguide backward scattering might be due to symmetric
waveguides while the backward-scattering in an asymmetric waveguides
can be avoided or reduced~\cite{PhysRevA.69.042303}.
Figure \ref{fig10}(b) shows the charge current density at the current-dip $L_{\rm CW} \simeq 110$~nm.
The incoming charge into the control-waveguide encounters the control-QD leading to electron back-scattering
in the control-waveguide, thus the net charge current is suppressed forming a current-dip.

We can say that the side-dots play a scattering role in the
waveguides in such way that the electrons partially get forward- and
backward-scattered in and into both waveguides.

\begin{figure}[htbq!]
      \includegraphics[width=0.45\textwidth,angle=0]{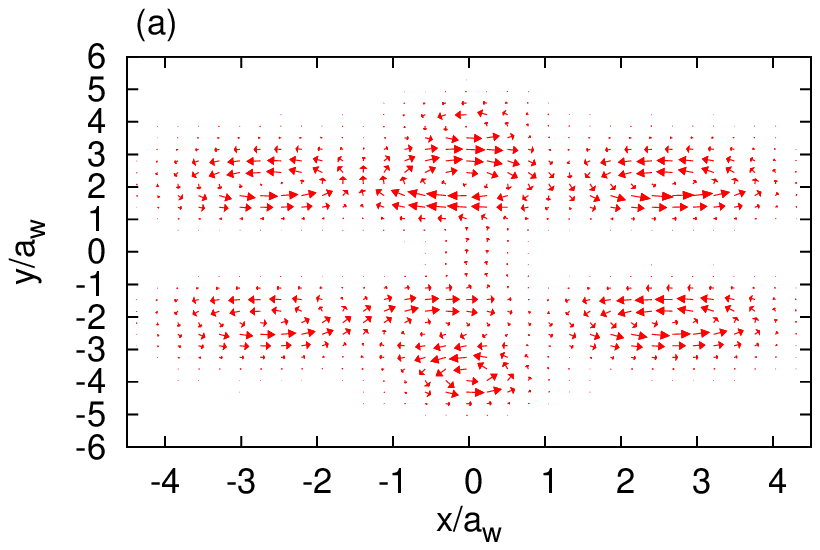}
      \includegraphics[width=0.45\textwidth,angle=0]{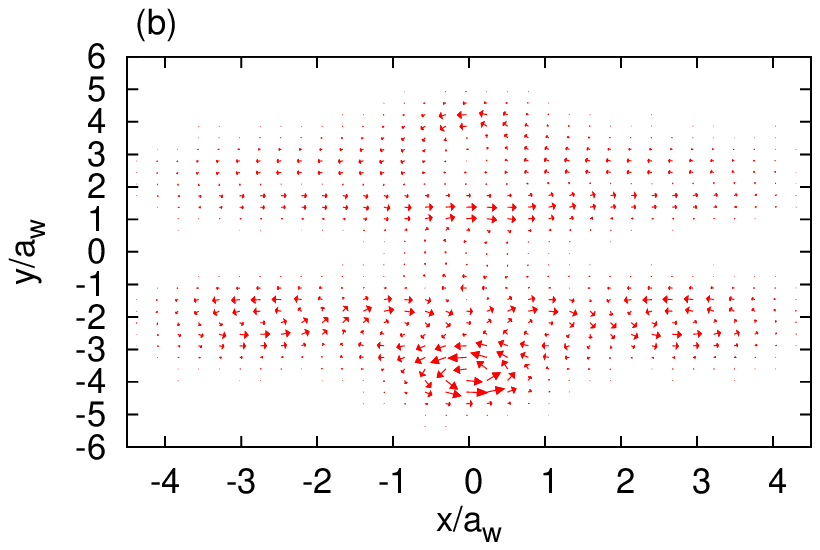}
      \caption{(Color online) Charge current density in the DQW at $t = 200$~ps
               in the presence of the Coulomb interaction
               in the peak $L_{\rm CW} \simeq40$~nm (a), and the dip $L_{\rm CW} \simeq 110$~nm (b)
               net charge current of \fig{fig08}(a) in the case of $B = 0.2$~T.
               Other parameters are $a_w=32.78$~nm, and $R_{\rm Dot} \approx 25$~nm.}
\label{fig11}
\end{figure}

By tuning the external magnetic field to a higher value, $B =
0.2$~T, the electrons manifest different dynamical motion in the
waveguides. In \fig{fig11} the charge current density is shown in
this stronger magnetic field $B = 0.2$~T at the CW length $L_{\rm
CW} \simeq 40$ (a) and $110$~nm (b). The graphs show circular
motion, edge states, in each waveguide.

In \fig{fig11}(a) is shown the current charge density at $L_{\rm CW} \simeq 40$~nm,
the electrons in the control-waveguide do not
totally switch to the target-waveguide, because of increased energy
level splitting at the resonance-energy in the presence of higher
magnetic field. In addition, charge is partially
localized in the control-QD as the effective magnetic length
approaches the radius of the quantum dot $a_w \simeq R_{\rm Dot}$.

In \fig{fig11} (b), the electrons are mostly localized in the
control-QD corresponding to the current-dip at $L_{\rm CW} \simeq
110$~nm. This electron localization is expected because at a larger CW
length the splitting of energy levels increases and leads to
suppression the inter-waveguide scattering. Therefore, the electrons
move along the control-waveguide and get localized in the control-QD
at this higher magnetic field.

\section{Concluding Remarks}\label{Sec:IV}

We have presented a model for a window coupled double
quantum waveguide in an external perpendicular magnetic field. The
DQW is weakly connected to two leads with different chemical
potentials in which a non-Markovian master equation is utilized to
investigate coherent switching in electron transport between the
waveguides for the implementation of quantum logic
gates.

A comparison between the charge current and the left and right
currents in Figures \ref{fig06} and \ref{fig08} at the time point
in time $t=200$~ps shows that the system has not completely reached a steady
state yet, but the sought after function of the DQW is already present
in the late transition regime. 
This is important. The exact steady state takes a long time to
reach due to a low rate of charging for 2 and 3 electron states.
Valuable time can be gained by operation of a device in the 
transition regime.

By tuning the length of the CW, we have demonstrated two important
physical characteristics of the waveguide system: resonant
energy-levels and oscillations in the net charge current. The
resonant energy levels indicate a strong coupling between the DQW
and the oscillations denote that the resonance conditions are
governed by the length of the CW.

In the case of an ideal non-Coulomb-interacting window-coupled 
DQW system, the charge current density splits equally between 
the waveguides at resonant energies, therefore, the net charge 
current reaches it's maximum value and
the waveguide system works as a $\sqrt{\rm NOT}$-operation qubit. In
the presence of strong magnetic field, the inter-waveguide backward
scattering is enhanced, and the efficiency of the logic gate
decreases.

This inter-waveguide forward scattering at resonant levels without
a magnetic field is not seen without including two-electrons states
in the model. It is strongly reduced by the Coulomb interaction that
lifts the 2ES out of the group of active states in the transport.  
This aspect can though be controlled by the exact system size
and design as the Coulomb blocking is reduced in a larger system,
or by using a higher bias window or different material.
Our models only includes sequential tunneling of electrons from the 
leads to the system, but the addition of higher order tunneling 
would only increase the importance of the many-electron structure 
and the Coulomb interaction on the operation of the system.

Two parallel waveguides or dots can be coupled in many distinct ways. 
Moldoveanu\cite{PhysRevB.82.085311} et al.\ attain it capacitively, 
while Zibold\cite{PhysRevB.76.195301} et al.\ fine tune the system by 
tunneling coupling. Both groups stress the strong dependence of the 
device on its exact geometry and the structure and correlations of 
many-electron states that are
active in the transport. Our investigation with a different coupling
scheme for the waveguides confirms this sensitive dependence.

In the presence of side QDs, more energy-levels of the two waveguides 
are brought into resonance which can lead to stronger coupling between 
them. Effectively, the dots can increase the density of states around the 
resonant transport states. 
The electrons from the control-waveguide switch to the target-waveguide
indicating a Not-operation qubit. But on the other hand, the
side-dots can also increase the inter-waveguide backward scattering
and reduce the qubit efficiency. In a stronger external magnetic
field, the electrons get localized in the dots as the effective
magnetic length is comparable to the radius of the side dot. In this
case, the geometry of the CW does not affect the electron transport
anymore.

The external magnetic field here is considered static, but one more
variable to influence the transport is the shape of the side QDs in
the waveguides.  We show that the varying of the CW length between
the waveguides can specify the type of the qubit logic gate to be
implemented by the system.

\begin{acknowledgments}
This work was financially supported by the Icelandic Research and
Instruments Funds, the Research Fund of the University of Iceland,
the Nordic High Performance Computing facility in Iceland, and the
National Science Council in Taiwan through Contracts No.\
NSC100-2112-M-239-001-MY3 and No.\ MOST103-2112-M-239-001-MY3.
\end{acknowledgments}

%
\bibliographystyle{apsrev4-1}

%
%
\end{document}